\documentclass[fleqn,10pt]{wlscirep}
\usepackage[utf8]{inputenc}
\usepackage[T1]{fontenc}
\usepackage{wrapfig}
\usepackage{gensymb}
\usepackage{sidecap}

\title{Enhancing spatial hearing with cochlear implants: exploring the role of AI, multimodal interaction and perceptual training.}

\author[1*]{Lorenzo Picinali}
\author[2]{Robert Baumgartner}
\author[3]{Valerie Gaveau}
\author[4, 5, 6]{Antonino Greco}
\author[7]{Stefanie Liebe}
\author[8]{Paul Oomen}
\author[4, 5, 6]{Christoph Braun}
\affil[1]{Audio Experience Design, Imperial College London, UK}
\affil[2]{Acoustics Research Institute, Austrian Academy of Sciences, Vienna, Austria}
\affil[3]{Centre de Recherche en Neurosciences de Lyon Inserm, Lyon, France}
\affil[4]{Department of Neural Dynamics and Magnetoencephalography, Hertie Institute for Clinical Brain Research, University of Tübingen, Germany}
\affil[5]{Centre for Integrative Neuroscience, University of Tübingen, Germany}
\affil[6]{MEG Center, University of Tübingen, Germany}
\affil[7]{University of Tübingen, Germany}
\affil[6]{NEMO Labs Nonprofit Kft., Budapest, Hungary}

\affil[*]{l.picinali@imperial.ac.uk}


\begin{abstract}

Cochlear implants (CIs) have been developed to the point where they can restore hearing and speech understanding in a large proportion of patients. Although spatial hearing is central to controlling and directing attention and to enabling speech understanding in noisy environments, it has been largely neglected in the past. We propose here a multi-disciplinary research framework in which physicians, psychologists and engineers collaborate to improve spatial hearing for CI users.

\end{abstract}

\begin{document}

\flushbottom
\maketitle
\thispagestyle{empty}

\section*{Introduction}
Hearing loss affects approximately 466 million people worldwide, requiring extensive medical care \cite{world2021world}. It also frequently leads to a significant decline in quality of life, resulting in social isolation, depression \cite{shield2019hearing}, and negatively impacting economic independence \cite{nordvik2018generic}. In addition to these adverse psychosocial and economical effects, hearing loss increases the risk of neurodegenerative diseases  such as dementia \cite{livingston2017dementia}. Thus, hearing loss has become an urgent health-related and economical problem, exacerbated by the increasing aging populations across all European regions. 

Based on decades of innovative advancements in biomedical technology, nowadays hearing can be at least partially restored for many hearing-impaired or deaf patients. Miniaturized hearing aids (HAs) and cochlear implants (CIs) capture acoustic signals and transmit the processed signals to the auditory system. While hearing aids transmit the recorded and amplified sound to the eardrum, CIs process the acoustic signals in real-time and generate electrical pulses that directly stimulate the auditory nerve. Essentially, CIs function as 'digital ears', representing the most advanced and successful solution to profound hearing loss and deafness. However, while HAs and CIs improve hearing for many patients in simple acoustic scenarios (e.g., talking to someone face-to-face), they are limited in naturalistic acoustic situations characterised by target sounds embedded in different, simultaneously active noise sources (e.g., a crowd of people in a party \cite{niparko2010spoken}).

In real-world scenarios, it is crucial to separate relevant sound sources from background noise, and spatial hearing (SH) plays a vital role in this process. Spatial hearing is essential for spatial orientation and navigation \cite{wallach1940role, brimijoin2012undirected, llado2024predicting}, and for directing attention to important sounds \cite{fritz2007auditory}, such as a looming object \cite{ignatiadis_cortical_2024}. This ability is particularly evident in the 'cocktail party effect' \cite{cherry1953some, lewald2016brain}, which describes the capacity to focus on a specific sound source, like a particular speaker, amidst a multitude of other sounds, such as the ambient noise of a party or following one speaker in a video conference. The ability to localize sound extends beyond auditory capabilities and impacts other cognitive functions, such as maintaining postural balance \cite{zhong2013relationship}, making it a crucial issue for aging individuals.

In normal hearing (NH) humans, sound localisation relies on binaural spatial cues such as interaural time difference (ITD) and interaural level difference (ILD) and on monaural cues such as spectral distortions due to the filtering effect of the pinnae, head, neck and shoulders \cite{blauert1997spatial}. However, whether and how SH can be achieved for CI patients remains largely unknown. Up to now, aspects of SH have largely been neglected in diagnostics and rehabilitation of patients with profound hearing loss, as well as in the development of HAs and CIs. 

Insufficient spatial hearing abilities following CI implantation \cite{coudert2022spatial} might have various causes, which can be broadly divided into two categories. 

The first one is related to the electrical signals delivered by CI to stimulate the cochlear nerve, which differ significantly from natural physiological conditions. In SH with CIs, spatial cues are extracted from the sound input using directional microphones and specialized filtering strategies. However, due to electronic sound processing and the limited number of electrodes stimulating the cochlear nerve, the temporal dynamics and frequency resolution of the signals transmitted to the auditory system differ from those in NH individuals. Furthermore, in bilateral CI users, the two processors are still operated in isolation from each other, resulting in two independent, uncoordinated auditory streams reaching the brain \cite{van2004exploring, grantham2007horizontal, seeber2008localization, dorman2014interaural, francart2015effect}. 

The second category  is connected with the rehabilitation following CI implantation, which often lacks sufficient training for spatial hearing. Prolonged deafness can cause neuroplastic changes in the brain, potentially impairing the ability to process spatial auditory information after CI implantation \cite{kral2016neurocognitive, heimler2014revisiting}. Despite these maladaptive changes, the assessment of sound localisation skills is not routinely included in medical examinations, and no systematic rehabilitation strategy for spatial hearing has been established.

To tackle this problem, there is a need to improve hearing restoration by combining the high sensitivity of machine learning approaches with higher processing capabilities of next-generation CIs, and employing innovative rehabilitation techniques following implantation. In this position paper, a series of research endeavours is proposed to tackle this matter in a coordinated manner, and at the same time to train the next generation of engineers and audiologists.

\section{Objectives}
Similarly to what has been suggested in the virtual audio domain by \cite{picinali2023system}, SH could be significantly improved by: (a) improving CI technologies (system-to-user adaptation), and (b) implementing rehabilitation procedures for SH (user-to-system adaptation) that foster auditory re-mapping and brain plasticity. 
The outcomes of both improvement strategies can be evaluated through assessment of psychophysical changes of SH. 
To obtain a mechanistic understanding of the interventions' working principles, our proposed approach consists of merging existing initiatives, sharing and improving models of the auditory system’s functions (e.g. ‘Auditory Modelling Toolbox’ \cite{majdak_amt_2022}). As part of these endeavours, neurophysiological processes underlying SH will be investigated using electroencephalography (EEG) and corresponding psychophysical measurements, and integrated into a deep neural network model simulating SH in CI-users. 

Specific objectives are:

\begin{enumerate}
    \item To investigate the manifold causes for altered and impaired SH at different levels of the auditory pathway in CI patients, in order to develop optimal therapeutic interventions. 
    
    \item To improve CI technology, fine-tuning the binaural timings and intensity levels of the pulses stimulating the auditory nerve to provide binaural cues of SH to CI patients. 
    
    \item Implementation and testing of novel behavioral training protocols to improve SH through three lines of research.
    \begin{enumerate}
        \item Investigating the benefits of a multisensory training of SH in CI users, with combined auditory-tactile and audio-visual stimuli. 
        \item Probing the role of head and body movements while actively exploring the acoustic environment. 
        \item  Employing training environments based on auditory and visual-auditory virtual realities (VR) and augmented realities (AR) to create enriched training environments that are optimally suited to the performance level of CI users. 
    \end{enumerate}
\end{enumerate}

\section{Proposed Workplan}
In order to identify key elements of SH and thus define the optimal stimulation conditions for its rehabilitation, implementing different strategies for the presentation of spatial sounds and their localisation using psychophysical approaches and comparing their effect on SH is fundamental. Spatial sounds will be presented to subjects either by (1) fixed or variable loudspeakers configurations in space \cite{toole2006loudspeakers, shah2011calibration}; (2) VR presenting sounds through earphones taking head position into account \cite{picinali2022sonicom}; or (3) 4DSOUND technology using real-time motion and position tracking to account for free movement and interactive behaviour of subjects in space \cite{holleman2019}. Tracking the subject's head and body movements during a spatial listening task allows for the quantification of the importance of active sensing strategies compared to passive listening (e.g., when the head and the body are kept still, limiting audio-visual and tactile integration). 

A second approach is to optimize CI parameters and functionalities. Since the processing of acoustic information in CIs is faster than in NH, introducing a delay in the CI stimulation is assumed to be beneficial to adjust the timing of the binaural sensory input in both unilateral implantations and bimodal hearing, i. e. hearing with CI on one ear and a hearing aid on the other ear \cite{zirn2019reducing}. For bilateral CIs, SH will likely be improved by maintaining ITDs through synchronisation of the stimulation of both ears. The use of adaptive filters (AF) can enhance CI by compensating for subject- and environment-dependent deviations between CI output and input, as such catering to real-time personalisation of the CI stimulation. AF implementations have been shown to improve intelligibility in HA and CI \cite{blamey2005adaptive} and may further improve synchronisation and balance of frequency-dependent ITD and ILD in bilateral CIs.

Due to the distributed cerebral processing of spatial sound information, neurophysiological methods are particularly helpful for identifying the physiological causes of SH impairments in CI users  for designing optimal rehabilitation measures. For example, EEG recordings can be used to obtain insights into how different brain regions collaborate to generate an auditory-spatial percept. The excellent temporal resolution of the EEG signals allows for tracking (1) the processing of the spatial test sounds in the auditory system using event-related potentials (ERPs) \cite{luck2014introduction, winkler2022auditory}; (2) the oscillatory brain activity reflecting the dynamics of brain states \cite{deng_impoverished_2019}, and (3) functional brain networks involved in processing spatial sounds \cite{nolte2004identifying, stam2007phase, toth2019attention}. 

Despite recent advances in AI across various fields, its impact on hearing healthcare has been limited \cite{lesica2021harnessing}. This emphasizes the need for interdisciplinary efforts to develop and implement AI technologies that can address the complexities of auditory processing in healthy and hearing-impaired populations. Recently, the process of sound localization has been successfully modeled by training a deep recurrent neural networks (RNN) performing spatial hearing tasks \cite{francl2022deep}. This approach can be extended by equipping the RNN with a biophysical CI-model \cite{hu2023model} during simulated intact, lesioned, and restored auditory processing in order to identify how SH is impaired in CI-users and to uncover potential targets for restorative rehabilitation. Advanced statistical methods including state-of-the art machine learning (ML) algorithms and DNN will be used to control spatial auditory training procedures. These procedures are based on different auditory scenes (walking in the city, adventure games, storytelling versus soundscapes, etc.) and various training strategies (training schedule, training with and without head, body and spatial movements, training with audio-visual, audio-tactile or audio-only inputs, etc.). Together with the knowledge accumulated through the creation of perceptual auditory models \cite{majdak_amt_2022}, ML will be used to predict the success of the training, considering patients’ individual preferences, personality traits and states as features. Satisfaction of patients with treatment will be assessed using standardised questionnaires used across all projects. 

\section{Projects}
EC funding, awarded through the CherISH (Cochlear Implants and Spatial Hearing) project, will allow us to engage in a series of projects organised in five distinct work packages:
\begin{enumerate}
\item
Computational neural network models will be used to describe auditory functions in NH and CI subjects. DNNs will model spatial-auditory processing and by comparing DNN activation patterns with EEG data, we will develop a neurophysiology-compatible model of SH information processing, enabling effective training simulations.
\item
Binaural CIs or one CI and a contralateral HA will be coupled with ultrafast interaural communication to synchronize devices, improving SH in binaural and bimodal CIs. Current binaural CI systems operate independently, making it difficult to use and learn SH cues. We will design a coordinated gain and synchronisation system to improve SH, evaluated before and after additional rehabilitative training.
\item
SH relearning will be facilitated by associating sounds with spatially congruent tactile or visual cues. SH training through auditory stimuli and non-auditory spatial cues requires an automated system that identifies sound origins and translates them into tactile or visual cues. We will develop an autonomous audio-tactile trainer that uses EEG to tailor tasks based on neurophysiological impairments.
\item
The active exploration of soundscapes via head and body movements will be investigated. Virtual reality helps to train hearing with CI, but the impact of auditory and visual approximations on success is unclear. We will will compare real sounds, loudspeakers, and headphone-based rendering, to assess conditions for optimal training outcomes. 
\item
Individual training programs will be tailored based on personality traits and SH performance. Training success depends on factors like motivation, resilience, and social inclusion. Using machine learning, we will model relationships between these factors and training outcomes, designing personalized training procedures.
\end{enumerate}

\section{Potential impact}
The proposed research has the potential to impact several domains and communities.

\subsection{Scientific impact}
‘Health throughout the life course’, ‘personalised medicine’, and 'tools, technologies and digital solutions for health and care' have been identified by the EU’s Health Research and Innovation Plan as relevant topics. Research within CherISH has the potential to make an important contribution to these areas by investigating efficient rehabilitation approaches and advanced technologies to improve spatial hearing of patients with profound hearing loss. With the expansion of CIs to include the processing of spatial-acoustic information, we have identified a pivotal development direction. The perception of a sound in space involves mapping of sensory features to spatial coordinates and heavily relies on learning and experience. Our innovative approach to neural plasticity of the human brain combined with advanced technology has the potential to enable the restoration of SH. To this end, modern AI-based technologies are used to model SH at different levels of processing and will serve as a test case of how complex operations such as SH can be studied and modeled in healthy as well as CI-based audition. 

\subsection{Educational impact} The proposed research will be carried out in parallel with a structured, high-quality and interdisciplinary doctoral training programme. Educational topics range from basic (physiology, psychology) to applied sciences (medicine, engineering) and are delivered through diverse learning measures like lectures, seminars, hands-on training and practice in regularly held educational workshops. The program will prepare candidates for scientific but also private sector careers by offering secondments with industry partners, providing insights into various professional practices and ensuring interdisciplinary knowledge transfer between science and industry. 

\subsection{Societal impact}
By producing highly trained and qualified professionals in the field of CI research, delivering innovative technology (AI-based sound processing, AR and VR for spatial sound) and providing CI users with newly developed hearing therapies, this research will aim at reducing the health and social impact of hearing loss at large.  Hearing loss is often associated with individual isolation, exclusion from professional life, social distancing, and the development of dementia \cite{world2021world,livingston2017dementia}. Improved spatial hearing in CI users will promote patients’ health, their personal independence and quality of life, as well as support their economic autonomy. 

\section{About the CherISH consortium}

The CherISH project is coordinated by Christoph Braun at the University of Tübingen (Germany), and includes partners from France (University Claude Bernard Lyon), Austria (Austrian Academy of Sciences and the University of Vienna), Belgium (Catholic University of Leuven and Cochlear Benelux NV), Hungary (Nemo Labs Nonprofit Kft, Budapest University of Technology and Economics and TTK Research Center for Natural Sciences), Italy (University of Trento), and UK (Imperial College London and Reactify Music LLP). More information about CherISH, as well as regular updates, can be found in the project website (\url{https://cherish-network.eu/}).

\section{Conclusions}

Cochlear Implants are probably one of the most successful examples of neuroprosthesis, and have been implanted successfully on hundreds of thousands patients worldwide. However, it is clear from both literature and clinical practice that individuals using bilateral cochlear implants are not fulfilling their potential due to problems with poor sound localisation and associated speech-in-noise perception. The CherISH project is aiming at addressing this matter by carrying out novel research in the domain, and at the same time by training the next generation of researchers, therapists and engineers to further tackle this and other related challenges. 

\section*{Author Contributions}
CB is the CherISH project coordinator, and is therefore the main contributor to the scientific parts of the paper. LP, RB, VG, AG, SL, PO and CB, coordinated by LP, contributed equally to the writing of the paper.

\section*{Competing interests}

The authors report there are no competing interests to declare.

\section*{Acknowledgements}

This work is funded by the European Union’s Horizon 2020 framework program for research and innovation under the Marie Sklodowska-Curie Grant Agreement No: 101120054.

\bibliography{mybib}

\end{document}